\documentclass[aps,journal=prb,amsmath,amssymb,unsortedaddress,noeprint,twocolumn]{revtex4-2}
\usepackage{graphicx}
\usepackage{epstopdf}
\usepackage{bmpsize}
\usepackage{amsmath,amsfonts,amsthm,bm}
\usepackage{tikz}
\usetikzlibrary{decorations.pathreplacing}
\usepackage[e]{esvect}
\usepackage{subcaption}
\usepackage[bookmarks=false]{hyperref}
\hypersetup{colorlinks=true, citecolor=blue, urlcolor=blue, linkcolor=blue}

\usepackage{stackengine,scalerel}
\usepackage{calc}
\newlength\shlength
\newcommand\xshlongvec[2][0]{\ThisStyle{\setlength\shlength{#1\LMpt}%
  \stackengine{-5.6\LMpt}{$\SavedStyle#2$}{\smash{$\kern\shlength%
    \stackengine{\dimexpr 1.3pt+6.25\LMpt}{$\SavedStyle\mathchar"017E$}%
      {\rule{\widthof{$\SavedStyle#2$}}{\dimexpr.1pt+.5\LMpt}\kern.4\LMpt}{O}{r}{F}{F}{L}\kern-\shlength$}}%
      {O}{c}{F}{T}{S}}}

\makeatletter
\renewcommand{\maketag@@@}[1]{\hbox{\m@th\normalsize\normalfont#1}}%
\makeatother

\newcommand{\bra}[1]{\ensuremath{\langle{#1}|\,}}
\newcommand{\ket}[1]{\ensuremath{\,|{#1}\rangle}}

\graphicspath{{figs/}}

\begin{document}

\author{Nicholas S. Davis}
\affiliation{College of Science and Engineering, James Cook University, Townsville, QLD, 4811, Australia }
\author{Samuel L. Rudge}
\address{Physikalisches Institut, Albert-Ludwigs Universit\"at Freiburg, Freiburg, 79100, Germany}
\author{Daniel S. Kosov}
\affiliation{College of Science and Engineering, James Cook University, Townsville, QLD, 4811, Australia }

\title{Electronic statistics-on-demand: bunching, anti-bunching, positive and negative correlations in a  molecular spin-valve}

\begin{abstract}
One of the long-standing goals of quantum transport is to use the noise, rather than the average current, for information processing. However, achieving this requires on-demand control of quantum fluctuations in the electric current. In this paper, we demonstrate theoretically that transport through a molecular spin-valve provides access to many different statistics of electron tunneling events. Simply by changing highly tunable parameters, such as electrode spin-polarization, magnetization angle, and voltage, one is able to switch between Poisson behavior, bunching and anti-bunching of electron tunnelings, and positive and negative temporal correlations. The molecular spin-valve is modeled by a single spin-degenerate molecular orbital with local electronic repulsion coupled to two ferromagnetic leads with magnetization orientations allowed to rotate relative to each other. The electron transport is described via Born-Markov master equation and fluctuations are studied with higher-order waiting time distributions. For highly magnetized parallel-aligned electrodes, we find that strong positive temporal correlations emerge in the voltage range where the second transport channel is partially open. These are caused by a spin-induced electron-bunching, which does not manifest in the stationary current alone.
\end{abstract}

\maketitle

\section{Introduction}\label{sec: Introduction}

The current generation of electronic technology has progressed rapidly through constant, iterative improvements to compactness. With smaller components, devices occupy less space within a circuit and dissipate less energy, allowing the construction of densely-packed circuits with improved computational power \cite{Hutchby2014}. This miniaturization process, however, is soon approaching the molecular size limit of a few nanometers. On this scale, the well-understood deterministic nature of electron transport is overturned by the probabilistic nature of quantum effects, opening new possibilities for information processing and corresponding research avenues. Among the most promising is the field of molecular spintronics \cite{Sanvito2011}, which explores how electron spin can be used in application to molecular-sized devices, as opposed to using just the electron charge \cite{Yakout2020}.

A molecular spintronic device typically consists of a molecule connecting two macroscopic electrodes.  The transition from electronics to spintronics demands that there be a level of magnetization within the system; some components must possess a dominant spin-orientation. Although one can achieve this through nonmagnetic leads and a single-molecule magnet \cite{Bogani2008,Zu2014}, in this paper, we focus exclusively on spin-valves created from a nonmagnetic molecule coupled to two ferromagnetic electrodes \cite{Sothmann2014,Guo2019,Braun2004,Prezioso2011,Tang2018,Hueso2007}. Such devices, commonly called magnetic tunnel junctions, have been experimentally realized with quantum dots \cite{Chye2002,Bernand-Mantel2006,Deshmukh2002,Hofstetter2010}, carbon nanotubes \cite{Feuillet-Palma2010,Aurich2010,Jensen2004,Jensen2005,Liu2006}, and even C$_{60}$ molecules \cite{Yoshida2013,Pasupathy2004}. 

The main idea of a molecular spin-valve is to control the flow of electrons via the magnetic orientation of the electrodes. Through magnetic properties alone, the valve can shift between high and low conductance states by rotation of the electrodes \cite{Konig2003,Braun2004}. Beyond this basic function, however, spin-valves also produce numerous regimes with interesting effects that, combined with the size advantage of a molecular device, make these devices preferable over conventional electronics. Two such effects that have already been demonstrated prototypically are spin-blockade induced negative differential resistance (NDR) \cite{Braun2004,Zhang2018} and spin filtering \cite{Zu2014,Wu2019}. Spin filtering describes a situation in which a particular spin orientation is preferentially allowed to flow through the system, creating spin-polarized current. The utility of spin filtering is expansive as it can be used to transfer large magnetic moments through a system, thus providing a means of flipping magnetic states, known as spin crossover \cite{Senthil2017}. 

\begin{figure}
	\centering
	\includegraphics[width=1 \linewidth]{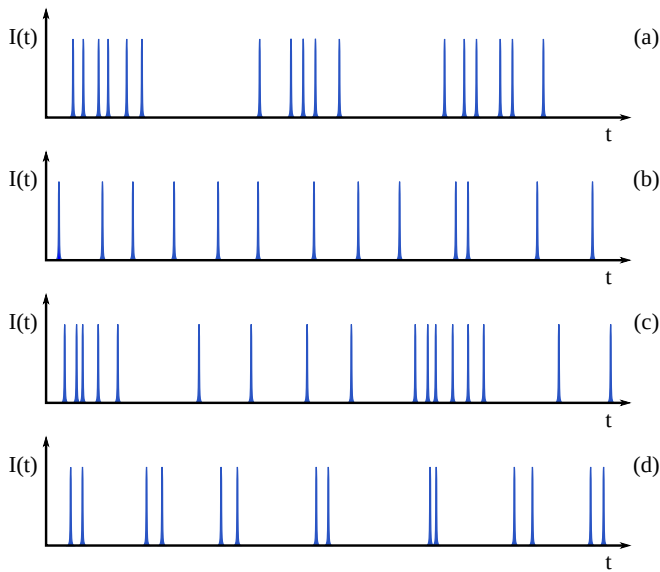}
	\caption{Examples of different statistics for electronic transport. Each peak represent electron detection event in the drain. In addition to the standard Poisson process, one can also observe: (a) electron bunching (super-Poisson statistics), (b) electron anti-bunching (sub-Poisson statistics), (c) positive temporal correlations, and (d) negative temporal correlations. All these different statistics can be observed in a molecular spin-valve.}
	\label{fig:1}
\end{figure}

In this paper, we show that, additionally to these interesting effects, a molecular spin-valve provides the possibility of controlling quantum current fluctuations on a single electron level. By adjusting highly tunable parameters, one can switch between different statistics of electron transport: Poisson, electron bunching (super-Poisson statistics),  electron anti-bunching (sub-Poisson statistics), and positive and negative temporal correlations, all of which are shown in Fig.(\ref{fig:1}). We explore these different fluctuation statistics using the waiting time distribution (WTD). 

While a relatively recent addition to the analysis of charge fluctuations in open quantum systems \cite{Brandes2008}, waiting time theory has developed rapidly in the last $10$ years. Consequently, it has been used to investigate a wide variety of transport scenarios, such as tunneling through molecules with electron-electron \cite{Ptaszynski2017,Brandes2008,Rudge2016b,Rudge2018} and electron-phonon \cite{Kosov2017b,Kosov2018a,Koch2006,Rudge2019a} interactions, telegraphic switching \cite{Rudge2019c}, double \cite{Ptaszynski2017a,Welack2009} and triple \cite{Engelhardt2019,Rudge2020} quantum dots, superconducting junctions \cite{Albert2016,Chevallier2016,Mi2018,Walldorf2018,Rajabi2013}, coherent conductors \cite{Dasenbrook2015,Haack2014,Albert2012}, non-Markovian transport \cite{Stegmann2020,Thomas2013}, periodically driven transport \cite{Albert2011,Albert2014}, and transport in the transient regime \cite{Tang2014a,Tang2014,Tang2017,SeoaneSouto2015}. As opposed to the full counting statistics (FCS), which is the most prevalent method for analyzing charge fluctuations, the WTD provides information on transport at short timescales, particularly via correlations between successive waiting times \cite{Ptaszynski2017,Albert2011,Budini2011,Ptaszynski2018,Albert2014,Rudge2019b}. Although there exist finite-frequency and time-dependent FCS capable of analysis at similar timescales \cite{Marcos2010,Ubbelohde2012,Stegmann2015,Stegmann2016}, as well as discussions of correlation behavior with current cumulants \cite{Sanchez2007,Sanchez2008a,Sanchez2008b}, most FCS treatments work in the long-time or zero-frequency limit \cite{Stegmann2017}. Measuring waiting times experimentally is a difficult task, requiring sophisticated single-electron detection methods \cite{Jenei2019,Singh2019,Ranni2021,Brange2021}; however, there are several novel proposals to measure waiting times, such as using a quantum `clock' \cite{Dasenbrook2016} and matrix product states \cite{Haack2015}.

In the context of quantum dot spin-valves, finite-frequency FCS have previously been used to analyze how the interplay between Larmor precession and a spin blockade affect the current noise \cite{Braun2006,Sothmann2010}, while the zero-frequency noise has been used to investigate the spin torque \cite{Virtanen2017} and dynamical blockade \cite{Cottet2004a,Cottet2004b,Lindebaum2009}. The WTD and time-dependent FCS have also been previously used to analyze the coherent dynamics present in this system, finding that Larmor precession causes coherent oscillations in the WTD and time-dependent charge cumulants \cite{Sothmann2014,Stegmann2018}. Several groups have also worked with spin-dependent fluctuation statistics, rather than just using the electron charge. Tang et al., for example, have recently developed the theory of spin-resolved waiting times in a quantum dot spin-valve \cite{Tang2018}, while S\'{a}nchez et al. have developed a correlation coefficient for the spin-resolved current \cite{Sanchez2008c}.  Our analysis, however, focuses on correlations between successive waiting times, and we find that, surprisingly, the strongest correlations appear when the coherent dynamics of the system are diminished.

The paper is organized as follows. Section \ref{sec: Theory} describes the theory; the model Hamiltonian, master equations, and the WTD, along with its main statistical tools, are introduced. Results are presented and discussed in Section \ref{sec: Results}, which discusses the average current, a measure of the noise, and temporal correlations. Section \ref{sec: Conclusions} presents the conclusions, while the explicit form of the Liouvillian and jump superoperators are given in Appendix \ref{app: Spin-resolved self-energies} and \ref{app: Spin-valve BMME in Liouville space}. 

We use natural units throughout the paper: $\hbar = e = 1$.
 \begin{figure}
\begin{center}
\includegraphics[width=1 \linewidth]{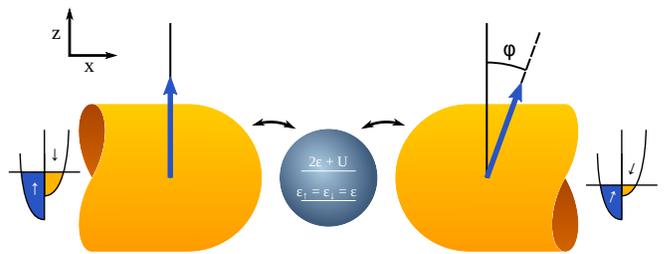}
\caption{Schematic of a molecular spin-valve, which consists of an Anderson model coupled to ferromagnetic electrodes. The spin quantization axis of the source electrode is aligned with the $z$-axis, whereas the drain electrode is allowed to rotate along the $x-z$ plane, enclosing an angle $\phi$ with the $z-$axis.}
\label{fig:2}
\end{center}
\end{figure}
\section{Theory}\label{sec: Theory}

\subsection{Model}

To model the spin-valve, we couple a molecule to two ferromagnetic leads, which has the total Hamiltonian
\begin{align}
H_{\text{tot}} & = H_{\text{mol}} + H_{\text{el}} + V.
\end{align}
Here, $H_{\text{mol}}$ is the molecular Hamiltonian, $H_{\text{el}}$ is the Hamiltonian of the source and drain electrodes, and $V = V_{S} + V_{D}$ enables tunneling between these two subsystems.

For the molecule, we use an Anderson model, with Hamiltonian
\begin{align}
H_{\text{mol}} & = \sum_{\sigma}\varepsilon d^{\dag}_{\sigma} d^{}_{\sigma} + U d^{\dag}_{\uparrow} d^{}_{\uparrow} d^{\dag}_{\downarrow}d^{}_{\downarrow},
\end{align}
where $\sigma \in \{\uparrow,\downarrow\}$ denotes spin in the molecule, $\varepsilon = \varepsilon_{\uparrow} = \varepsilon_{\downarrow}$ is a spin-degenerate molecular orbital energy, and $U$ is the Coulomb interaction between two electrons occupying the same orbital. We choose the spin quantization axis of the Anderson model to be along the $z$-direction and refer to it as the `global' spin orientation; the `local' quantization direction of the ferromagnetic electrodes will be taken relative to this. 

The source and drain ferromagnetic electrodes are reservoirs of non-interacting electrons
\begin{align}
H_{\text{el}} & = \sum_{\alpha \in \{S,D\}} \sum_{\mathbf{k}_{\alpha},\sigma_{\alpha}}\varepsilon^{}_{\mathbf{k}_{\alpha}  \sigma_{\alpha}}c_{\mathbf{k}_{\alpha} \sigma_{\alpha}}^{\dagger}c^{}_{\mathbf{k}_{\alpha}\sigma_{\alpha}},
\end{align}
where the subscript $\mathbf{k}_{\alpha}$ denotes the energy states of lead $\alpha \in \{S,D\}$. The electrodes are held at local thermal equilibrium with the same temperature, $T_{\alpha} = T$, but different chemical potentials, $\mu_{\alpha}$. In all calculations, we adjust the chemical potentials symmetrically around the Fermi level, which is kept at zero: $\mu_{S} = V_{SD}/2 = -\mu_{D}$, where $V_{SD} = \mu_{S} - \mu_{D}$ is the source-drain bias voltage. 

To explicitly include the ferromagnetic nature of the leads, we apply an asymmetry to the spin-dependent density of states, $\rho_{\alpha\sigma_{\alpha}}(\omega)$, for majority, $\sigma_{\alpha} = + $, and minority, $\sigma_{\alpha} = - $, spins; see Fig.(\ref{fig:2}). The spin-polarization, 
\begin{align}
p_\alpha & = \frac{\rho_{\alpha +} - \rho_{\alpha -} }{\rho_{\alpha +} + \rho_{\alpha -}}, 
\end{align}
thus describes the level of magnetization of lead $\alpha$. The `majority spin' and `minority spin' terminology is used here to reflect the fact that the electrodes may have their spin quantized along different axes to the global spin orientation in the molecule. In our treatment, we always choose the quantization axis of electron spin in the source electrode to be along the $z$-axis, thus keeping it aligned with the molecule so that $\sigma_{S} \in \{\uparrow,\downarrow\}$. The drain electrode, however, is allowed have spin quantized in any direction in the $x$-$z$ plane, which encloses an angle $\phi$ with the $z$-axis. 

The different magnetization angles are also reflected in the source and drain tunneling Hamiltonians, $V = V_{S} + V_{D}$, which take the form 
\begin{align}
V_{S} & = \sum_{\mathbf{k}_{S}}  t^{}_{S}\left(c^{\dag}_{\mathbf{k}_{S}+}d^{}_{\uparrow} + c^{\dag}_{\mathbf{k}_{S}-}d^{}_{\downarrow}\right) + \text{h.c.}, \label{eq: V_S}\\
V_{D} & = \sum_{\mathbf{k}_{D}}t^{}_{D}\left\{\left[c^{\dag}_{\mathbf{k}_{D}+}\cos\left(\frac{\phi}{2}\right) - c^{\dag}_{\mathbf{k}_{D}-}\sin\left(\frac{\phi}{2}\right)\right]d^{}_{\uparrow}\right. \nonumber \\
& \:\:\:\:\:\:  \left. + \left[c^{\dag}_{\mathbf{k}_{D}+}\sin\left(\frac{\phi}{2}\right) + c^{\dag}_{\mathbf{k}_{D}-}\cos\left(\frac{\phi}{2}\right)\right]d_{\downarrow}\right\} + \text{h.c.} .\label{eq: V_D}
\end{align}
In Eqs.\eqref{eq: V_S}-\eqref{eq: V_D}, we have assumed that the tunneling matrix elements, $t_{\alpha}$, are real and independent of electrode energy and spin. Since we further assume that the spin-dependent density of states in each electrode is constant, $\rho_{\alpha,\sigma_{\alpha}}(\omega) = \rho_{\alpha,\sigma_{\alpha}}$, the tunneling matrix elements are related to the spin-dependent electrode-system coupling strength via $\Gamma_{\alpha\sigma_{\alpha}} = 2\pi|t_{\alpha}|^{2}\rho_{\alpha\sigma_{\alpha}}$, where we can define the corresponding electrode-dependent coupling strength and total coupling strength as $\Gamma_{\alpha} = \left(\Gamma_{\alpha,+} + \Gamma_{\alpha,-}\right)/2$ and $\Gamma = \left(\Gamma_{\alpha} + \Gamma_{\alpha}\right)/2$, respectively.

\subsection{Born-Markov master equation}\label{subsec: Born-Markov master equation}

Although the total dynamics are described by the total density matrix, $\rho_{\text{tot}}$, we actually only need the reduced density matrix for the molecular degrees of freedom, $\rho_{\text{mol}}$, since the electrodes are reservoirs of non-interacting electrons. The corresponding Hilbert space is spanned by the four orthonormal eigenstates of $H_{\text{mol}}$: the dot can be empty, $\ket{0}$, occupied by a single spin-up, $\ket{\uparrow}$, or spin-down, $\ket{\downarrow}$, electron, or two electrons of opposite spin, $\ket{2}$. The coupled elements of the reduced density matrix are
\begin{align}
\rho_{\text{mol}} & = \left[\begin{array}{c c c c} \rho_{00} & 0 & 0 & 0 \\ 0 & \rho_{\uparrow\uparrow} & \rho_{\uparrow\downarrow} & 0 \\ 0 & \rho_{\downarrow\uparrow} & \rho_{\downarrow\downarrow} & 0 \\ 0 & 0 & 0 & \rho_{22} \end{array} \right],
\end{align}
where most off-diagonal elements of $\rho_{\text{mol}}$ describe coherences between states of different charge occupancy, and so decouple from the transport. The two single electron coherences, $\rho_{\sigma\bar{\sigma}}$, are necessary in this treatment, as the spin accumulated within the system extends into the $x$ and $y$ directions. 

The reduced system density matrix time-evolves according to a generalized quantum master equation (GME). Although most previous treatments of the quantum dot spin-valve have used a real-time diagrammatic approach to calculate the required transition rates \cite{Braun2004,Braun2006,Sothmann2014}, in the weak-coupling limit a Born-Markov master equation (BMME) approach is equivalent. Here, the time-evolution of $\rho_{\text{mol}}$ amounts to solving the Liouville-von Neumann equation in the interaction picture, after tracing out the electrode degrees of freedom:
\begin{align}
\frac{d}{dt}\rho_{\text{mol},I}(t) & = \text{Tr}_{\text{el}} \left[V_{I}(t),\rho_{\text{tot},I}(t)\right], \label{eq: LvN interaction picture}
\end{align}
where operators in the interaction picture are $A_{I}(t) = e^{i\left(H_{\text{mol}} + H_{\text{el}}\right)t}Ae^{-i\left(H_{\text{mol}} + H_{\text{el}}\right)t}$.

Under the Born approximation, the influence of the Anderson-type molecule on the evolution of the electrodes is negligible and the total density matrix factorizes as $\rho_{\text{tot}} = \rho_{\text{mol}}\rho_{\text{el}}$. If one further assumes that correlations in the electrodes decay faster than the timescale of system dynamics, that is, that the transport is Markovian, then one can apply standard steps to Eq.\eqref{eq: LvN interaction picture} and obtain a Redfield-type master equation in the basis of eigenstates of $H_{\text{mol}}$ \cite{Breuer2002,Redfield1965}:
\begin{widetext}
\begin{align}
i\dot{\rho}^{}_{mn} & = \omega_{mn}\rho^{}_{mn} + \sum_{\alpha}\sum_{\sigma\sigma'}\sum_{kl}  \: \Sigma_{\alpha,\sigma\sigma'}^{>}(\omega_{lk}) \langle m|d^{\dag}_{\sigma}|k\rangle\langle k| d^{}_{\sigma'}|l\rangle \rho^{}_{ln} + \Sigma_{\alpha,\sigma\sigma'}^{<}(\omega_{kl})^{*} \langle m|d^{}_{\sigma}|k\rangle\langle k|d^{\dag}_{\sigma'}|l\rangle\rho^{}_{ln} \nonumber \\
& \:\:\:\: + \Sigma_{\alpha,\sigma\sigma'}^{<}(\omega_{nl}) \langle m |d^{\dag}_{\sigma}|k\rangle \rho^{}_{kl}\langle l |d^{}_{\sigma'}|n\rangle + \Sigma_{\alpha,\sigma\sigma'}^{>}(\omega_{ln})^{*} \langle m|d^{}_{\sigma}|k\rangle\rho^{}_{kl}\langle l |d^{\dag}_{\sigma'}|n\rangle - (m\leftrightarrow n,k \leftrightarrow l)^{*}, \label{eq: ME final}
\end{align}
\end{widetext}
where the last term indicates that the remaining four terms of the master equation are obtained by switching the positions of $\bra{m}$ and $\ket{n}$, as well as $\bra{k}$ and $\ket{l}$, then taking the Hermitian conjugate of the result. The exact forms of the lesser, $\Sigma_{\alpha,\sigma\sigma'}^{<}(\omega)$, and greater, $\Sigma_{\alpha,\sigma\sigma'}^{>}(\omega)$, self-energies are in Appendix \ref{app: Spin-resolved self-energies}.

In Liouville space, Eq.\eqref{eq: ME final} becomes a time-local first-order differential equation,
\begin{align}
\dot{\boldsymbol{\rho}}(t) & =  \mathbf{L} \boldsymbol{\rho}(t),
\label{master-eq}
\end{align}
where $\mathbf{L}$ is the time-independent Liouvillian superoperator containing all system dynamics, which is also written explicitly in Appendix \ref{app: Spin-valve BMME in Liouville space}, and the system density matrix has been mapped to a vector in Liouville space, $\rho_{\text{mol}}(t) \rightarrow \boldsymbol{\rho}(t)$, with 
\begin{align}
\boldsymbol{\rho}(t) & = \left[\rho_{00},\rho_{\uparrow\uparrow},\rho_{\uparrow\downarrow},\rho_{\downarrow\uparrow},\rho_{\downarrow\downarrow},\rho_{22}\right]^{T}.
\end{align}
The original formulation of Braun et al. \cite{Braun2004}, which many subsequent treatments also use \cite{Braun2006,Sothmann2010,Sothmann2014}, rewrites the GME as a master equation for the system charge states coupled to Bloch-like equations for the $x$, $y$, and $z$ components of the dot spin: $\boldsymbol{\rho} = \left[\rho_{00},\rho_{11},\rho_{22},S_{x},S_{y},S_{z}\right]^{T}$, with $\rho_{11} = \rho_{\uparrow\uparrow} + \rho_{\downarrow\downarrow}$ and 
\begin{align}\left\{S_{x},S_{y},S_{z}\right\} = \left\{\frac{\rho_{\uparrow\downarrow} + \rho_{\downarrow\uparrow}}{2},i\frac{\rho_{\uparrow\downarrow} - \rho_{\downarrow\uparrow}}{2},\frac{\rho_{\uparrow\uparrow} - \rho_{\downarrow\downarrow}}{2}\right\}.
\end{align}
Although our master equation in Eq.\eqref{eq: ME final} and the corresponding Liouvillian in Eqs.\eqref{eq: Liouvillian defn}-\eqref{eq: matrix D} are not in this form, we note that they describe exactly the same dynamics as long as the molecule-electrode coupling is weak: $\Gamma \ll k_{B}T$, which all of our results conform to.

All calculations are performed in the nonequilibrium steady state, $\bar{\boldsymbol{\rho}}$, which is the null vector of the full Liouvillian:
\begin{align}
\mathbf{L} \bar{\boldsymbol{\rho}} & = 0. \label{ss}
\end{align}

\subsection{WTD}\label{subsec: WTD}

Let us begin to monitor time delays between successive quantum tunnelings in the nonequilibrium steady state. Throughout the paper, we are solely concerned with tunnelings to the drain, which are contained in the quantum jump operator $\mathbf{J}$. To identify the exact form of $\mathbf{J}$, one could formally rewrite Eq.\eqref{eq: ME final} as an $n$-resolved master equation \cite{Li2005b} and include all terms associated with $\rho^{(n-1)}(t)$ \cite{Rudge2020}. However, one could also just directly identify those terms in Eq.\eqref{eq: ME final} that lower the system's charge occupation by $1$ via a tunneling to the drain, in this case including all single particle coherences; this is the approach we take and the resulting quantum jump operator is shown explicitly in Appendix \ref{app: Spin-valve BMME in Liouville space}. 

The WTD for $n$ consecutive  waiting times, $w_n(\tau_n, \dots,\tau_1) $, is defined as the joint probability distribution that,  given an initial tunneling to the drain, the next electron waits a time $\tau_{1}$ before tunneling to the drain, the electron after that then waits a time $\tau_{2}$ before tunneling to the drain, and so on up to the $n$th successive electron:
 \begin{align}
w_{n}(\tau_{n}, \dots,\tau_1) & =  \frac{\text{Tr} \left[\prod\limits_{m=1}^{n}\left(\mathbf{J}e^{\left(\mathbf{L}-\mathbf{J}\right)\tau_{m}}\right)\mathbf{J}\bar{\boldsymbol{\rho}}\right]}{\text{Tr}\left[\boldsymbol{J}\bar{\boldsymbol{\rho}}\right] },
\label{wtd_n}
\end{align}
where one can obtain Eq.\eqref{wtd_n} either from an extension of Brandes' phenomenological formula \cite{Brandes2008} or the idle-time approach defined by van Kampen \cite{VanKampen1981,Rudge2019b}. 

To compute expectation values and analyze the fluctuations, we introduce the moment-generating function (MGF) for the joint WTD,
\begin{align}
K_{n}(\vec{\chi}) & = \int_{0}^{\infty}  d \vec{\tau}  \: e^{i\vec{\chi} \cdot \vec{\tau}} w_{n}(\vec{\tau}), \label{eq: MGF}
\end{align}
which contains vectors of waiting times, $\vec{\tau}  = \left(\tau_{n}, \dots,\tau_{1}\right)$, and counting fields, $\vec{\chi}  = \left(\chi_{n},\dots,\chi_{1}\right)$. From the MGF in Eq.\eqref{eq: MGF}, we can calculate the moments of the joint WTD as
\begin{align}
\langle \tau_{n}^{k_{n}} \dots \tau_{1}^{k_{1}} \rangle & = \frac{\partial^{k_{n}}}{\partial (i\chi_{n})^{k_{n}}}\dots\frac{\partial^{k_{1}}}{\partial (i\chi_{n})^{k_{1}}}K_{n}(\vec{\chi})\Big|_{\chi_{n}=\dots\chi_{1}=0};
\end{align}
this is a relatively simple task given that the integral over $\vec{\tau}$ can be computed analytically:
\begin{align}
K_{n}(\vec\chi)= (-1)^{n}\frac{\text{Tr} \left[ \prod\limits_{m=1}^n \left(\mathbf{J}\left(\mathbf{L}-\mathbf{J}+i\chi_{m}\right)^{-1}\right)\mathbf{J}\bar{\boldsymbol{\rho}}\right]}{\text{Tr}\left[\mathbf{J}\bar{\boldsymbol{\rho}}\right]}.
\label{k2}
\end{align}
 
Our analysis will focus on the first and second moments of the first-order WTD, 
\begin{align}
\langle \tau^{k} \rangle & = \int_{0}^{\infty} d\tau_1  \; \tau^{k} w_{1}(\tau_{1}) \\
 & =  k! (-1)^{k+1} \frac{\text{Tr}\left[\mathbf{J}\left(\mathbf{L}-\mathbf{J}\right)^{-(k+1)}\mathbf{J}\bar{\boldsymbol{\rho}}\right]}{\text{Tr}\left[\mathbf{J}\bar{\boldsymbol{\rho}}\right]},
\end{align}
as well as the {   two-time correlations of the n-order WTDs, which are
\begin{widetext}
\begin{align}
\langle \tau_{i} \tau_{j} \rangle & = \int^{\infty}_{0} d\vec{\tau} \tau_{i}\tau_{j} w_{n}(\vec{\tau}) 
=
(-1)^{i-j-1}\frac{\text{Tr} \left[    \mathbf{J} \left(\mathbf{L} - \mathbf{J}\right)^{-2} \left( \mathbf{J} \left(\mathbf{L}-\mathbf{J}\right)^{-1}\right)^{i-j -1} \mathbf{J} \left(\mathbf{L} - \mathbf{J}\right)^{-2}\mathbf{J}\bar{\boldsymbol{\rho}}\right]}{\text{Tr}\left[\mathbf{J}\bar{\boldsymbol{\rho}}\right]}.
\end{align}

\end{widetext}
}

\subsubsection{WTD fluctuation statistics}

We now introduce the randomness parameter \cite{Ptaszynski2018,Rudge2019b},
\begin{align}
R=\frac{\langle\tau^{2}\rangle-\langle\tau\rangle^{2}}{\langle\tau\rangle^{2}}, \label{R}
\end{align}
to provide a measure of variance relative to the mean and characterize the Poissonian nature of the transport in the same manner as the widely used Fano factor, $F = \frac{\langle I^{2}\rangle - \langle I \rangle^{2}}{\langle I \rangle}$. If the counting of detected electrons is a Poissonian process, then $F = 1$, as the variance of a Poisson process is equal to its mean. Super-Poissonian noise, $F > 1$, therefore, indicates a variance greater than the mean, which is produced by electron bunching. The opposite scenario of sub-Poissonian noise, $F < 1$, then indicates electron `anti-bunching'. These processes are shown in Fig.(\ref{fig:1}). 

The associated WTD of a Poisson process is exponential: $w_{\text{Po}}(\tau) = \lambda e^{-\lambda \tau}$, with corresponding average $\langle \tau \rangle = 1/\lambda$ and variance $\langle \tau^{2} \rangle - \langle \tau \rangle^{2} =1/ \lambda^{2}$. As a result, the randomness parameter is also $R = 1$ for a Poisson process, with the same interpretations for super- and sub-Poissonian noise as well. 

{  
Correlations between successive tunneling events will be measured using a linear Pearson correlation coefficient, 
\begin{align}
PC_{\tau_{n},\tau_{1}} & = \frac{\langle  \tau_{n} \tau_{1} \rangle - \langle \tau \rangle^2 }{  \langle   \tau^2 \rangle- \langle   \tau \rangle^2 },
\label{pearson}
 \end{align}
where we are explicitly only considering the Pearson coefficient of the first, $\tau_{1}$, and last, $\tau_{n}$, waiting times in a sequence, as it contains nontrivial information about possible correlations in $w_{n}(\tau_{n},\dots, \tau_{1})$. Other two-time correlations can always be reduced to the averages performed over the lower order WTDs. For example, consider the second-order mean 
\begin{align}
\langle \tau_{n-k} \tau_l \rangle   
& = \int_{0}^{\infty} \prod\limits_{m=1}^{n} d \tau_m  \; \tau_{n-k}\tau_{l} w_{n}(\tau_{n},\dots, \tau_{1}) \\
& = \int_{0}^{\infty} \prod\limits_{m=l}^{n-k} d \tau_m  \; \tau_{n-k}\tau_{l} w_{n-l-k+1}(\tau_{n-k},\dots, \tau_{l}).
\end{align}

The Pearson correlation coefficient is widely used in statistics as a measure of the average linear correlation between two stochastic variables. In the waiting time context, a positive value, $1 > PC_{\tau_{n},\tau_{1}} > 0$, indicates that if the first waiting time, $\tau_{1}$, is longer (shorter) than average, then it is probable that the waiting time between the $n^{\text{th}}$ and $(n+1)^{\text{th}}$ electron tunnelings, $\tau_{n}$, will also be longer (shorter) than average. Negative correlations, $0 > PC_{\tau_{n},\tau_{1}} > -1$, in contrast, indicate that a longer (shorter) than average $\tau_{1}$ is likely to be followed by a shorter (longer) than average $\tau_{n}$.

We will also use a function that measures the correlation between each possible pair of the first two successive waiting times $\tau_{1}$ and $\tau_{2}$:
\begin{align}
C(\tau_{2},\tau_{1}) & = \frac{w_{2}(\tau_{2},\tau_{1})}{w_{1}(\tau_{2})w_{1}(\tau_{1})} - 1. \label{eq: correlation function}
\end{align}
$C(\tau_{2},\tau_{1})$ is interpreted similarly to the Pearson coefficient, although it provides a measure of the correlation between each pair of successive waiting times, rather than an average over the entire space of $\tau_{1}$ and $\tau_{2}$. 
}

\section{Results}\label{sec: Results}

\subsection{Stationary current}

In Ref.\cite{Braun2004}, Braun et al. present plots of the stationary current as a function of bias voltage in the same parameter regimes that we will probe with waiting times. We will not reproduce the plots here, which are included in the supplemental material, but will outline a brief explanation of the physics as a way of contextualizing our later results.
In the following, we will refer to three magnetization angles: the parallel case is when the leads magnetization is along the same axis, $\phi = 0$; the perpendicular case is when the drain is magnetized at an angle of $\phi = \frac{\pi}{2}$ to the source electrode; and the anti-parallel case corresponds to a relative magnetization angle of $\phi = \pi$. We also note that all discussions are for an equal spin-polarization of the leads: $p_{S} = p_{D} = p$. 

The parallel scenario shows, somewhat counter-intuitively, the same current profile as transport through a standard Anderson model with no lead magnetization. One might expect that, as the lead magnetization increases and tunneling is dominated by majority spin electrons, only one transport channel will be activated, even at high voltages when double occupancy is energetically allowed. However, the small percentage of minority spin electrons still play an active role in the transport; although they only rarely tunnel into the dot, they also only rarely tunnel out, spending a much greater time on the dot than majority spin electrons, thus creating a spin-blockade. Since $\langle I \rangle$ is measured in the long-time limit, and the original probability for a minority spin electron to tunnel in from the source is small, it cannot distinguish between these time-scales for majority and minority spins, and so `sees' the same transport as through an Anderson-type molecule coupled to nonmagnetic electrodes. 

The average current of anti-parallel leads, in contrast, displays a significant dependence on the strength of the magnetization within the leads. One still observes the double step behavior associated with two distinct transport channels, but for $p \rightarrow 1$ the current is severely diminished. Here, majority spins from the source electrode are treated as minority spins in the drain electrode, and vice versa, which means that total tunneling through the system, and thus the current, is highly suppressed. 

Besides this behavior, the most interesting current feature discussed by Braun et al. occurs in the perpendicular case, which displays NDR between $\varepsilon \leq V_{SD}/2 \leq \varepsilon + U$; that is, when a single transport channel is open. Braun et al. explain that, in this regime, increasing voltage tends to align the accumulated spin of the dot in the opposite direction to the magnetization in the drain, which strongly suppresses the current. The effect is modified by the inclusion of the virtual exchange field, which produces Larmor-like spin precession.

\subsection{Bunching and anti-bunching}

\begin{figure*}
\centering
\begin{subfigure}{0.32\textwidth}
\centering
\includegraphics[width=1\textwidth]{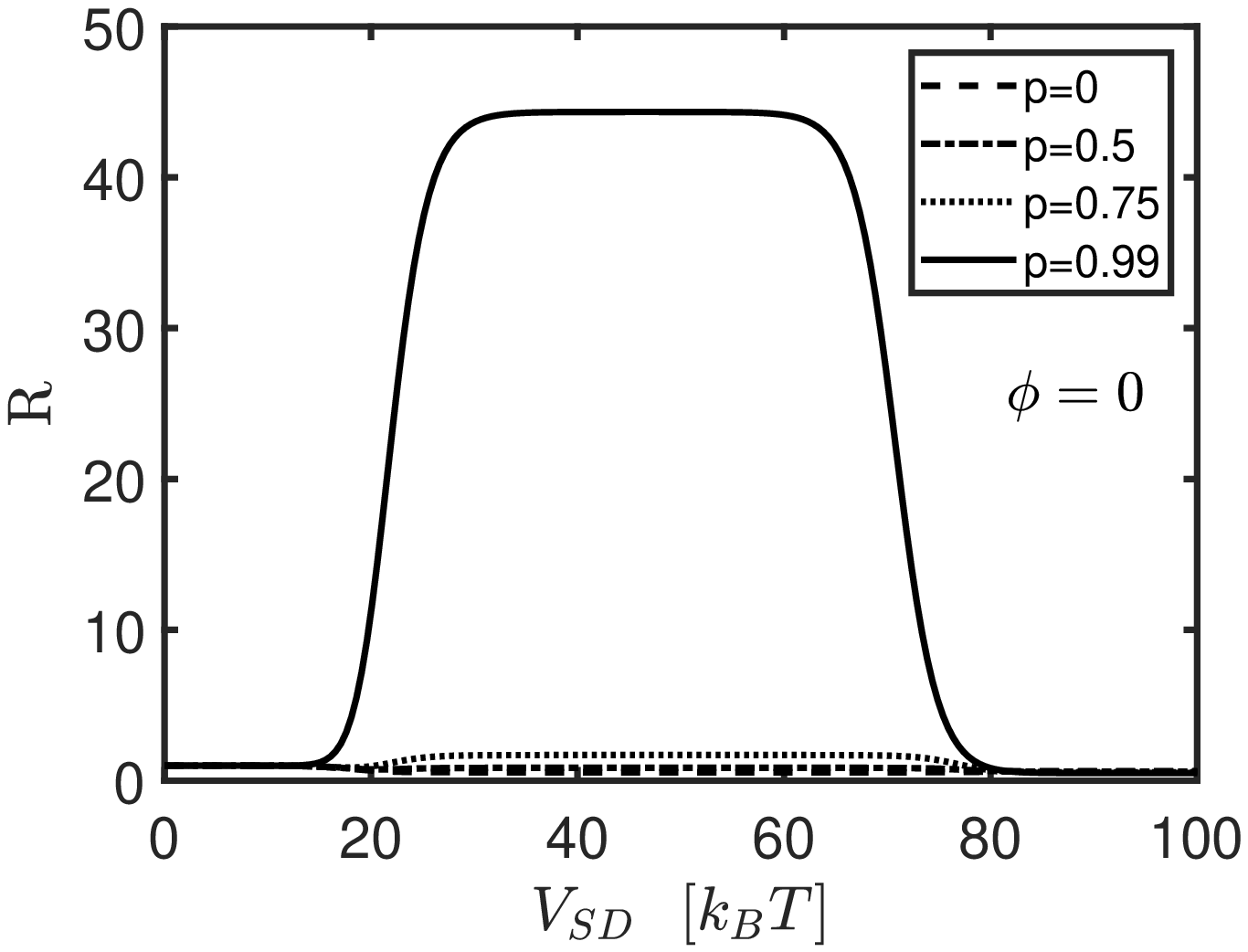}
\caption{}
\label{RV_parallel}
\end{subfigure}
\begin{subfigure}{0.32\textwidth}
\centering
\includegraphics[width=1\textwidth]{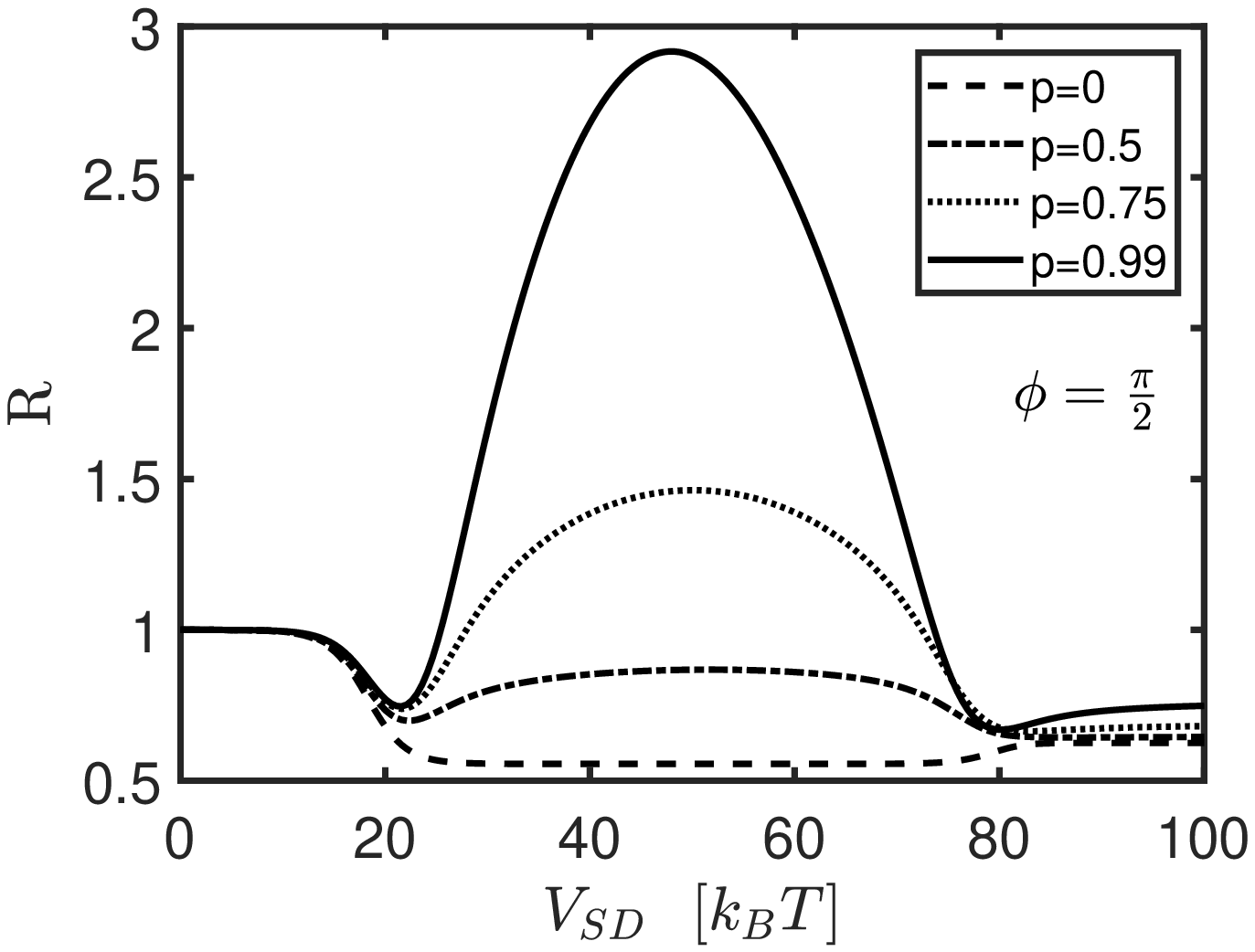}
\caption{}
\label{RV_perpendicular}
\end{subfigure}
\begin{subfigure}{0.32\textwidth}
\centering
\includegraphics[width=1\textwidth]{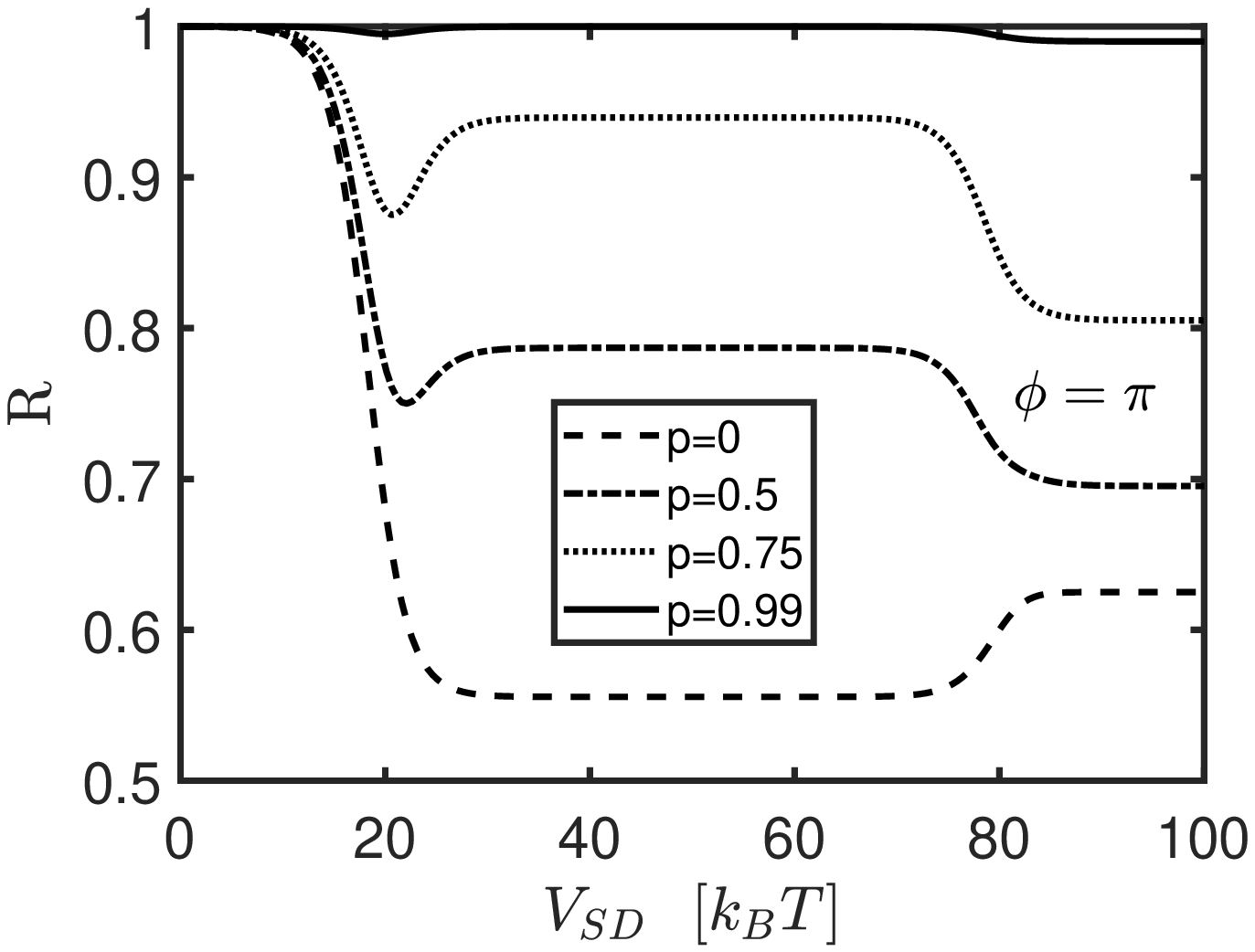}
\caption{}
\label{RV_anti-parallel}
\end{subfigure}
\caption{Randomness parameter, $R$, as a function of bias voltage for different spin-polarizations and relative magnetization angles: (a)  $\phi=0$, (b) $\phi=\frac{\pi}{2}$,  (c) $\phi=\pi$. Both electrodes have the same spin-polarization: $p_{S} = p_{D} = p$. Other parameters are $\varepsilon = 10 k_{B} T$,  $U = 30 k_B T$, $k_{B}T = 0.1\text{meV}$, $\gamma = 0.01\text{meV}$.}
\label{RV}
\end{figure*}

In this section, we will analyze the transport using the randomness parameter instead of the stationary current, as shown in Figs.(\ref{RV_parallel})-(\ref{RV_anti-parallel}). We note that we get similar results to that found by Braun et al. in Ref.\cite{Braun2006} for the Fano factor. First, as one can see in Fig.(\ref{RV_parallel}), parallel lead magnetizations and strong electrode spin-polarization produce highly super-Poissonian transport, as $R \gg 1$ when a single transport channel is open in $\varepsilon \leq V_{SD}/2 \leq \varepsilon + U$. This originates directly from the effect we discussed before; although transport is dominated by majority spin electrons, when minority spin electrons tunnel into the dot, they tend to stay there for a long time, due to a lack of available states in the drain. This evidently causes electron bunching in tunnelings to the drain, and the transport becomes strongly super-Poissonian. 

Fig.(\ref{RV_perpendicular}) also displays super-Poissonian transport, although the underlying physical reason now originates from the NDR discussed above. As electrons tunnel through the system, the accumulated spin tends to align in a direction anti-parallel to the drain magnetization, so that after a series of tunnelings eventually a spin-down electron becomes stuck in the dot: it cannot tunnel back to the source due to the bias voltage; no other electrons can tunnel in because double occupancy is energetically forbidden, and it cannot tunnel to the drain because there is a low density of states for that spin-polarization. Again, this produces electron bunching and $R > 1$, although we note that that magnitude is much reduced compared to Fig.(\ref{RV_parallel}). 

In the case of anti-parallel magnetization angles, shown in Fig.(\ref{RV_anti-parallel}), the transport is at all voltages sub-Poissonian. At $p = 0$, the $R$ is that of transport through a normal Anderson model, with behavior that is well-known in the literature \cite{Bagrets2003,Nazarov1996,Chen1992}. As the spin-polarization increases, however, electron tunneling through the system is almost completely suppressed and the time between tunnelings to the drain 
exponentially increases, pushing the transport towards Poissonian behavior. 

Overall, Figs.(\ref{RV_parallel})-(\ref{RV_anti-parallel}) demonstrate that a quantum dot spin-valve is able to produce statistics on demand; simply by changing the relative magnetization angle one is able to produce super-Poissonian, sub-Poissonian, and Poissonian noise.

\subsection{Temporal correlations}

\begin{figure*}
\centering
\begin{subfigure}{0.32\textwidth}
\centering
\includegraphics[width=1\textwidth]{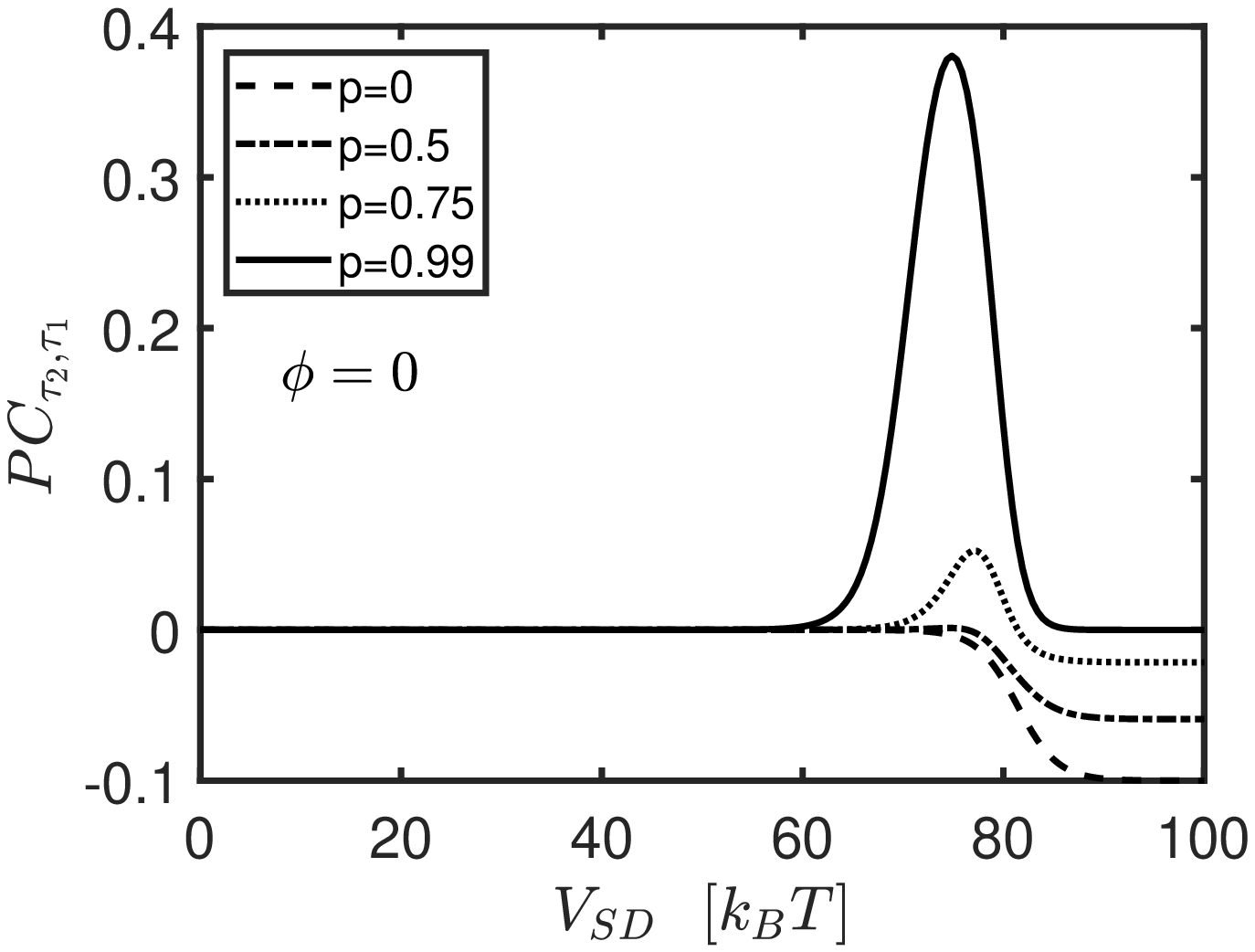}
\caption{}
\label{PV_parallel}
\end{subfigure}
\begin{subfigure}{0.32\textwidth}
\centering
\includegraphics[width=1\textwidth]{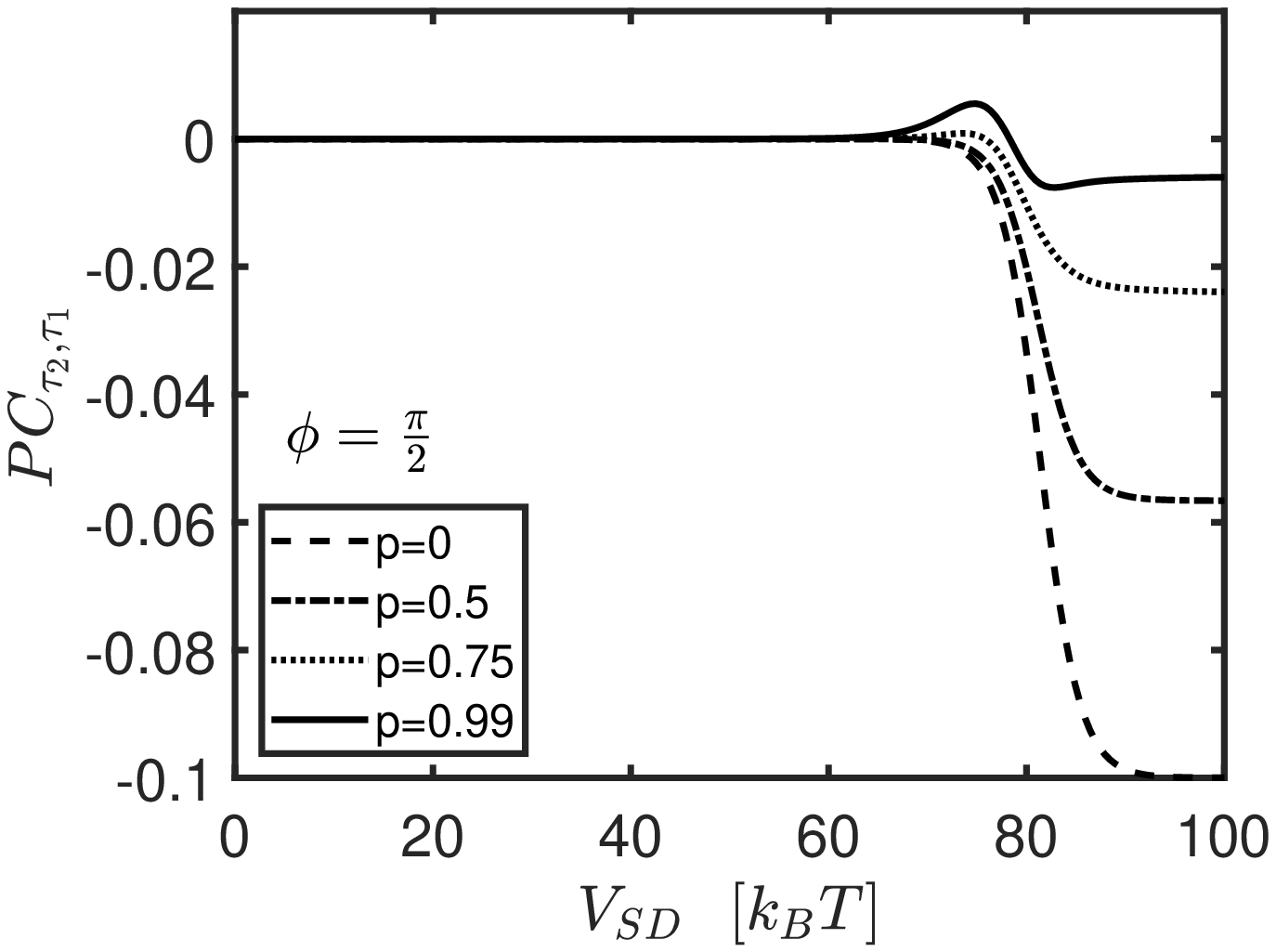}
\caption{}
\label{PV_perpendicular}
\end{subfigure}
\begin{subfigure}{0.32\textwidth}
\centering
\includegraphics[width=1\textwidth]{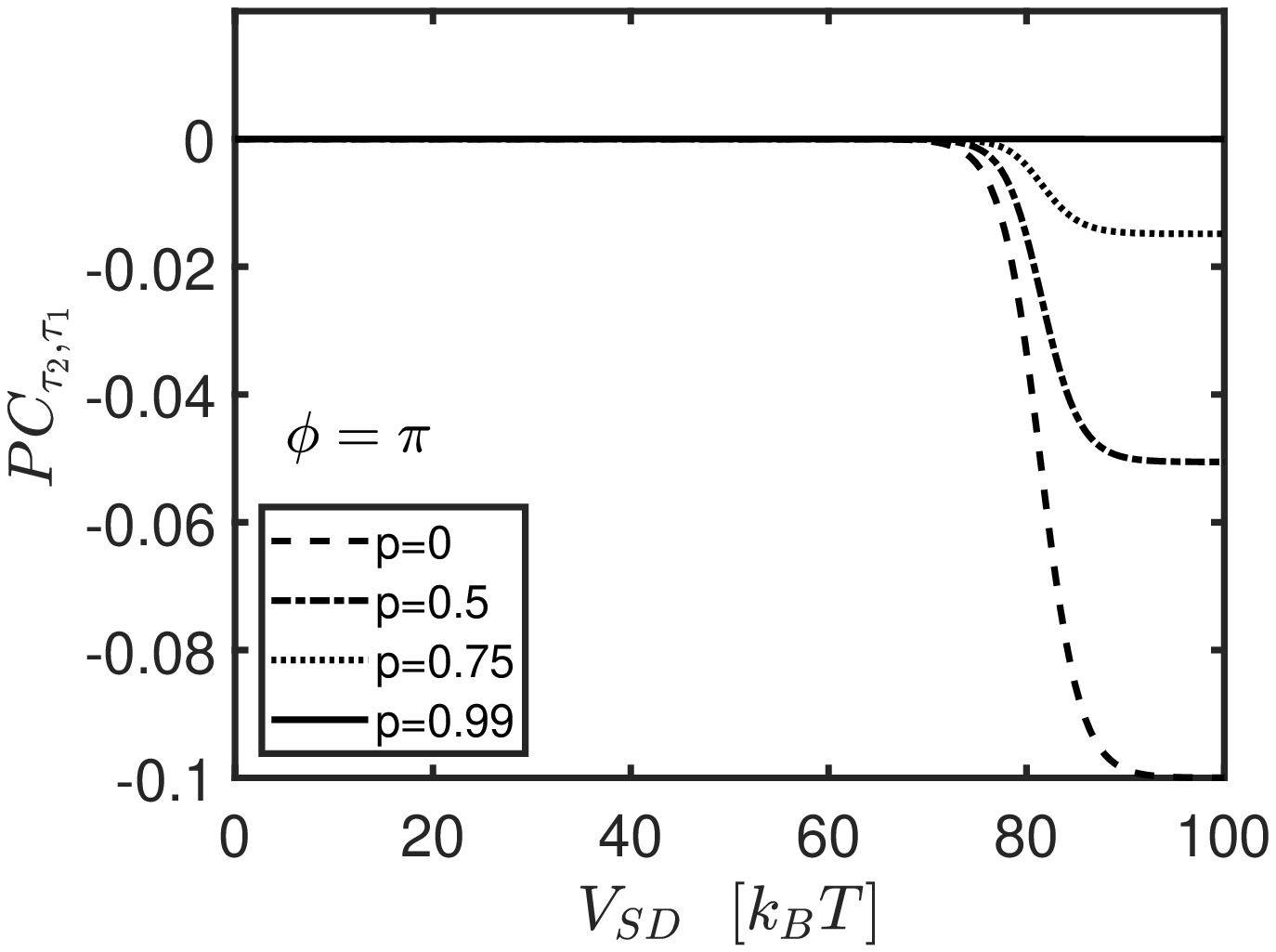}
\caption{}
\label{PV_anti-parallel}
\end{subfigure}
\caption{Pearson correlation coefficient, $PC_{\tau_{2},\tau_{1}}$, between the first two successive waiting times, $\tau_{1}$ and $\tau_{2}$, as a function of bias voltage for different spin-polarizations and relative magnetization angles: (a)  $\phi=0$, (b) $\phi=\frac{\pi}{2}$,  (c) $\phi=\pi$. All parameters are the same as in Figs.(\ref{RV_parallel})-(\ref{RV_anti-parallel}).}
\label{RV}
\end{figure*}

Beyond just the bunching and anti-bunching available in plots of the randomness parameter, the spin-valve also displays temporal correlations in certain regimes, which one can see in Figs.(\ref{PV_parallel})-(\ref{PV_anti-parallel}). 

{   At all magnetization angles and for small lead spin-polarization, one sees correlation behavior between the first two successive waiting times similar to that already reported for a standard Anderson model \cite{Ptaszynski2017,Rudge2018}.} At low voltages, when zero or one transport channel is open, successive waiting times are completely uncorrelated. In the double occupancy regime, however, the dot can fill with two electrons. A short initial waiting time, $\tau_{1}$, therefore, likely means that both electrons tunnel to the drain in quick succession. The dot then has to be filled with an electron from the source before another tunneling to the drain can occur, making the second waiting time, $\tau_{2}$, much longer; $\tau_{1}$ and $\tau_{2}$ become negatively correlated. The linear strength of this correlation is evidently not very large, as $\max\{|PC|\} = 0.1$, and it completely disappears for highly polarized leads. In the highly spin-polarized anti-parallel case, it is now unlikely that the dot will be doubly occupied and also unlikely that both of those electrons will tunnel to the drain in quick succession. In the perpendicular case, the reasoning is similar, but the magnetization angles are closer so spin-precession can somewhat mitigate the correlation suppression.

\begin{figure}
\centering
\includegraphics[width=0.7\columnwidth]{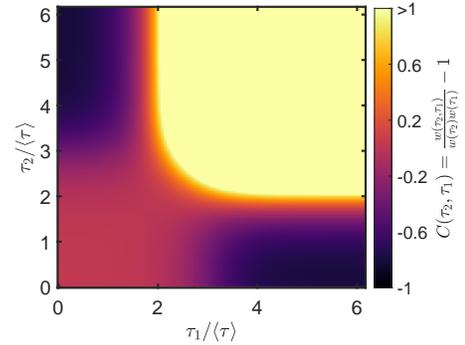}
\caption{Relative correlation between successive waiting times, $C(\tau_{2},\tau_{1})$, plotted for parallel magnetizations: $\phi = 0$. Spin-polarization within the leads is $p = 0.99$ and the voltage is $V_{SD} = 75k_{B}T$, corresponding to the peak in the positive correlations from Fig.(\ref{PV_parallel}). All other parameters are the same as in Figs.(\ref{RV_parallel})-(\ref{RV_anti-parallel}).}
\label{wtd2}
\end{figure}

{   In Fig.(\ref{PV_parallel}), however, one can see that, beside the already discussed correlation behavior at high voltages, parallel magnetization} and high lead spin-polarization produces strong positive correlations between successive waiting times. This occurs in the small voltage window corresponding to the second transport channel opening, as one can see by direct comparison with the stationary current. A more comprehensive picture of these correlations is shown in Fig.(\ref{wtd2}), which plots the two-time correlation function defined in Eq.\eqref{eq: correlation function} at the voltage corresponding to the Pearson coefficient maximum.

Here, one can see that the large Pearson coefficient originates from positive correlations between two \textit{long} waiting times: $\tau_{1},\tau_{2} > 3\langle \tau \rangle$. Although in the plot we have constricted the range to values between $ - 1 < C(\tau_{1},\tau_{2}) < 1$, so as to see the fine detail, the actual values go much higher. The correlation between two extremely long successive waiting times, for example, is $C(6\langle \tau \rangle,6\langle \tau \rangle) = 20$. Correspondingly, a short initial waiting time and a long secondary waiting time, or vice versa, are strongly negatively correlated.

\begin{figure*}
\centering
\begin{subfigure}{0.4\textwidth}
\centering
\includegraphics[width=1\textwidth]{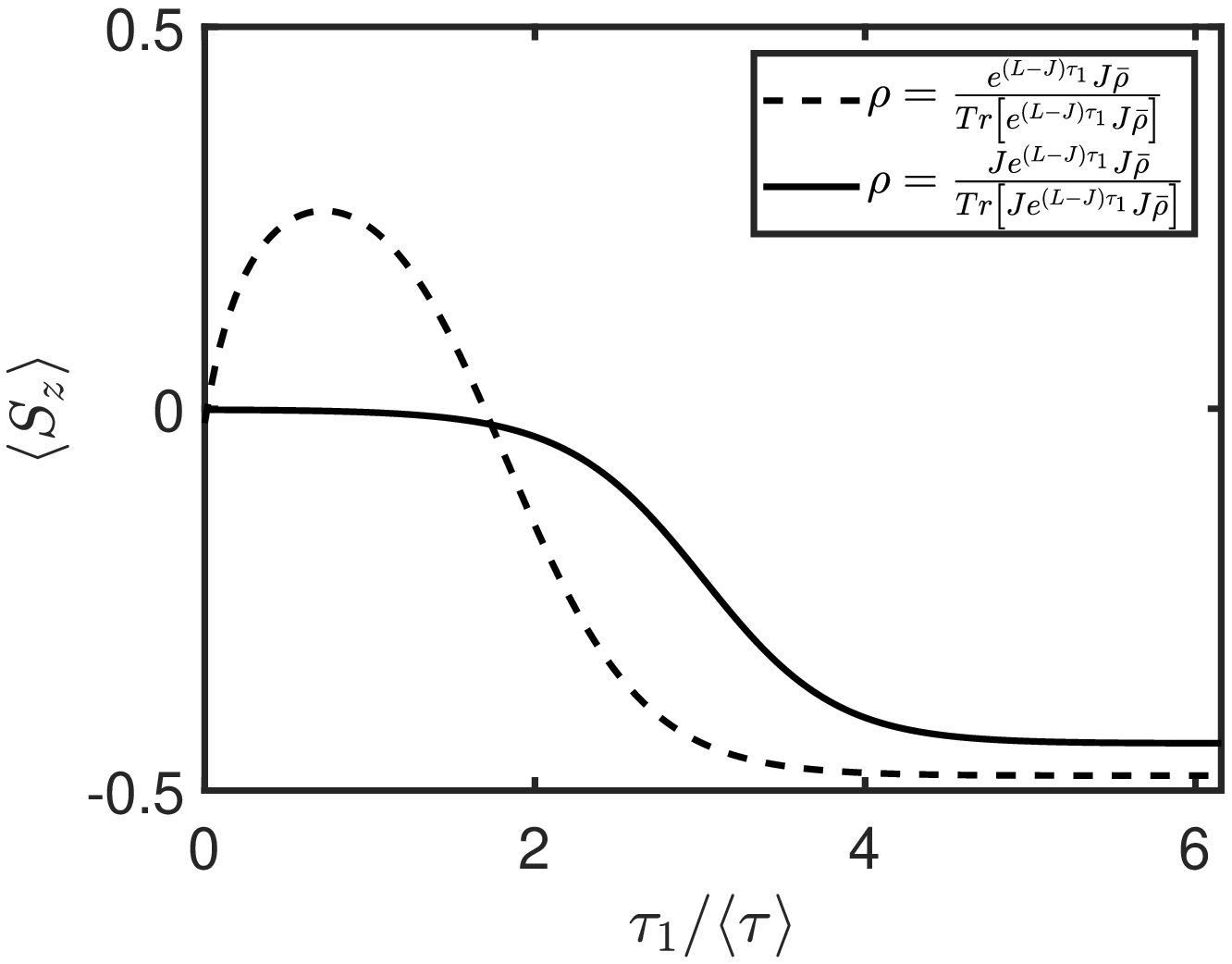}
\caption{}
\label{fig: 6a}
\end{subfigure}
\begin{subfigure}{0.4\textwidth}
\centering
\includegraphics[width=\columnwidth]{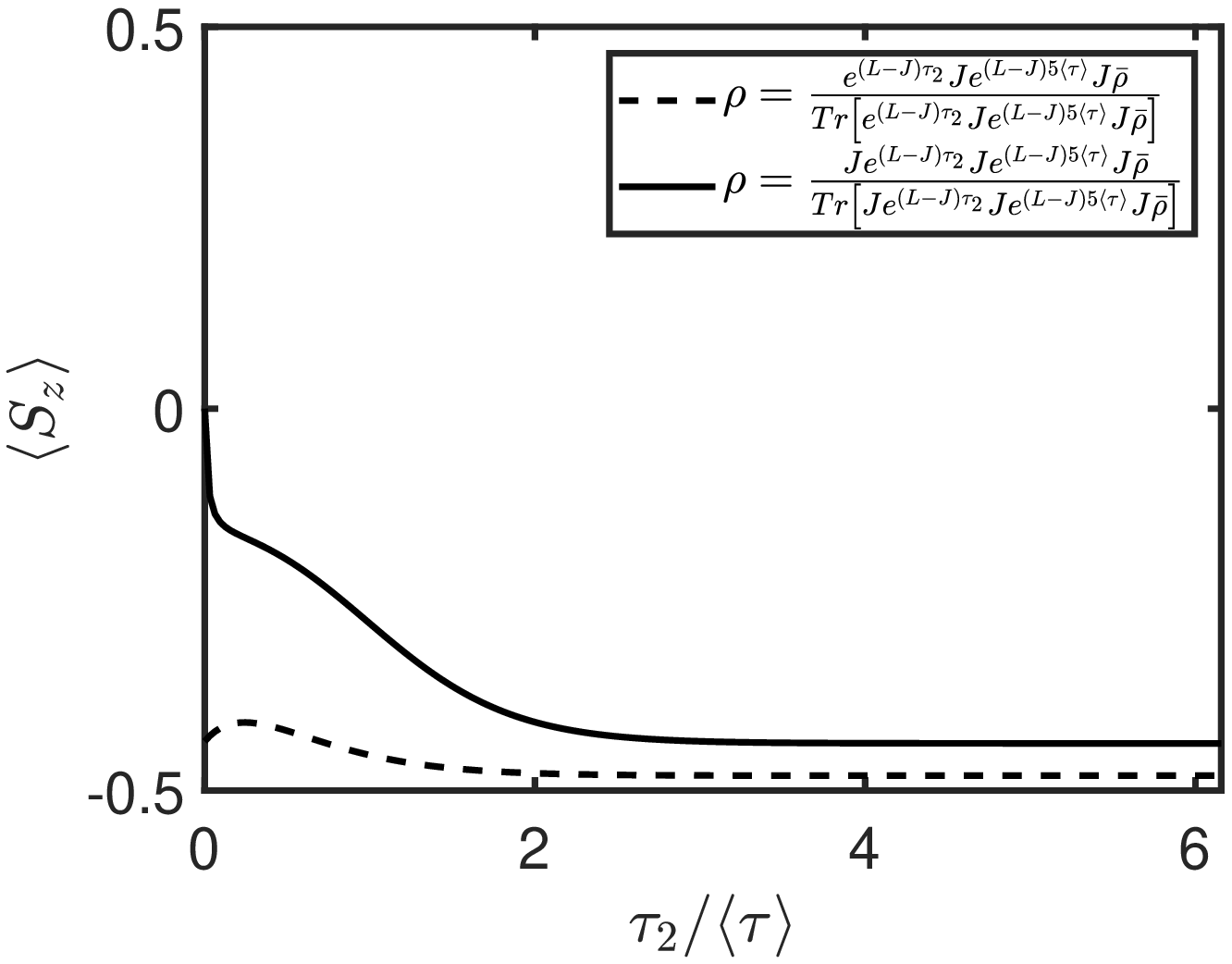}
\caption{}
\label{fig: 6b}
\end{subfigure}
\caption{$\langle S_{z} \rangle$, the $z$-component of the measurement-averaged spin, {   (a) as a function of the initial waiting time, $\tau_{1}$, and (b) as a function of the second waiting time, $\tau_{2}$, given a long initial waiting time: $\tau_{1}=5\langle\tau_{1}\rangle$.} All parameters are the same as in Fig.(\ref{wtd2}).}
\end{figure*}

{  
To understand the nature of these temporal correlations for pairs of long waiting times, consider Fig.(\ref{fig: 6a}). Here, we have plotted the $z$-component of the dot spin averaged over the quantum measurement projected density matrix, $\langle S_{z} \rangle = (\rho_{\uparrow\uparrow}-\rho_{\downarrow\downarrow})/2$, as a function of the initial waiting time, $\tau_{1}$. Note that, since the lead magnetizations are parallel, the $x$- and $y$-components of the average spin are zero. The dashed line corresponds to the projected density matrix of a system that starts in the stationary state, undergoes a jump to the drain, $\mathbf{J}$, and then evolves for some time without a jump to the drain, $e^{\left(\mathbf{L} - \mathbf{J}\right)\tau_{1}}$:
}
\begin{align}
\boldsymbol{\rho}(\tau_{1}) & = \frac{e^{(\mathbf{L}-\mathbf{J})\tau_{1}}\mathbf{J}\bar{\boldsymbol{\rho}}}{\text{Tr}\left[e^{(\mathbf{L}-\mathbf{J})\tau_{1}}\mathbf{J}\bar{\boldsymbol{\rho}}\right]}.
\end{align}
For $\tau_{1} < 2\langle \tau \rangle$, the accumulated spin in the dot is in the positive $z$ direction, so that it is likely to be occupied by majority spin electrons from the source. Fig.(\ref{fig: 6a}) shows, however, that when this majority spin electron tunnels out to the drain, modeled now with
\begin{align}
\boldsymbol{\rho}(\tau_{1}) & = \frac{\mathbf{J}e^{(\mathbf{L}-\mathbf{J})\tau_{1}}\mathbf{J}\bar{\boldsymbol{\rho}}}{\text{Tr}\left[\mathbf{J}e^{(\mathbf{L}-\mathbf{J})\tau_{1}}\mathbf{J}\bar{\boldsymbol{\rho}}\right]},
\end{align}
then the accumulated spin in the dot returns to zero and there is no preference for a tunneling of either spin, apart from that imposed by lead spin-polarization.

In contrast, if the initial waiting time is longer, $\tau_{1} > 2\langle \tau \rangle$, then the accumulated spin in the dot tends to be negative, and this accumulated spin remains after the next tunneling to the drain. {   Since this is within the voltage regime where the second transport channel is beginning to open, which one can see by direct comparison with $\langle I \rangle$, it is likely that this second jump to the drain is not due to the minority spin electron tunneling out, but rather a majority spin tunneling in from the source and out quickly to the drain. Consequently, the system remains likely to be occupied by a minority spin electron and this low probability mechanism is again required for the next jump, ensuring that $\tau_{2}$ is long as well.

Indeed, one can see this directly in Fig.(\ref{fig: 6b}), where we have plotted $\langle S_{z} \rangle$ during $\tau_{2}$, after setting $\tau_{1} = 5\langle \tau \rangle$. The dashed-line indicates the dot is highly likely to possess a minority spin during the $\tau_{2}$. At small $\tau_{2}$, $\langle S_{z} \rangle$ is close to zero, which demonstrates the relatively small chance that this jump was a minority spin electron leaving the system. Note that we have not shown the case of a short $\tau_{1}$, as the resulting plot is extremely similar to Fig.(\ref{fig: 6a}); the system empties itself with a majority spin tunneling to the drain, and so the dynamics in $\tau_{1}$ repeat.}

\begin{figure*}
\centering
\begin{subfigure}{0.4\textwidth}
\centering
\includegraphics[width=1\textwidth]{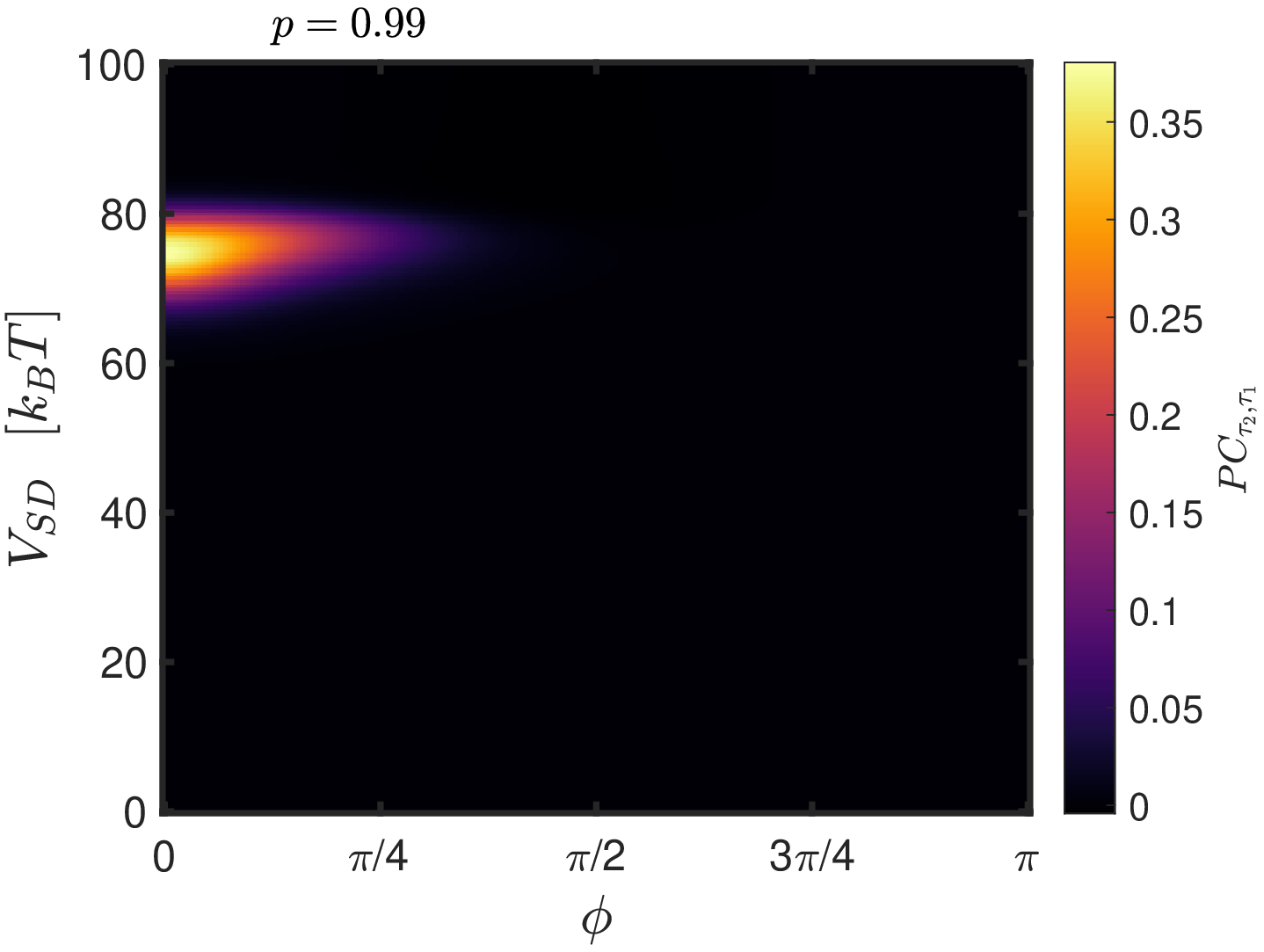}
\caption{}
\label{fig: 7a}
\end{subfigure}
\begin{subfigure}{0.4\textwidth}
\centering
\includegraphics[width=\columnwidth]{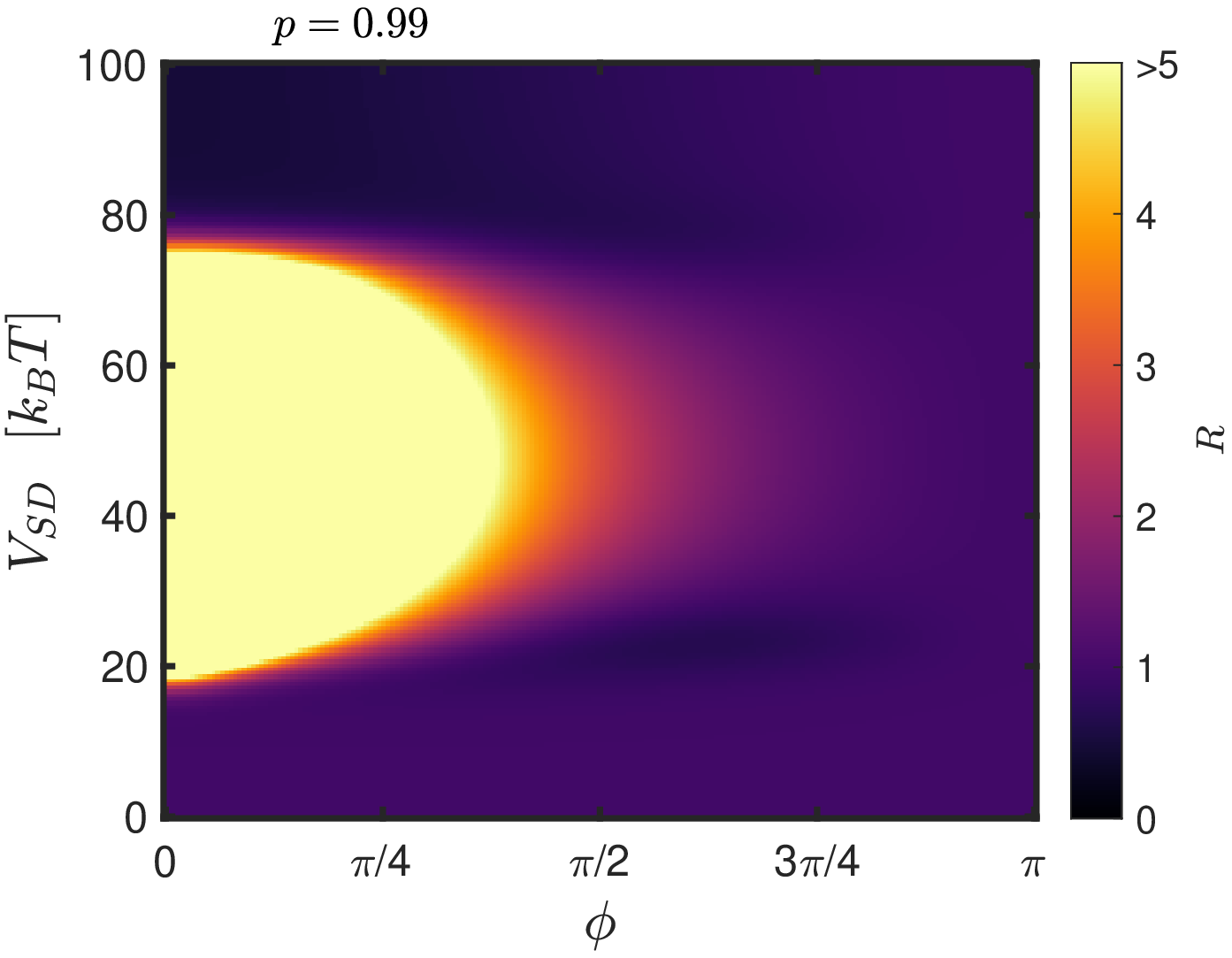}
\caption{}
\label{fig: 7b}
\end{subfigure}
\caption{Transport characteristics of the spin-valve across the relevant voltage range, $V_{SD}$, and half-rotation of the drain electrode, $\phi$. In (a) the successive waiting time correlation, $PC_{\tau_{2},\tau_{1}}$, is plotted, while (b) contains the randomness parameter, $R$. The spin polarization is $p=0.99$ and other parameters are the same as in Fig.(\ref{wtd2}).}
\label{fig: 7}
\end{figure*}

{   Evidently, these correlations are highly dependent on the leads being strongly polarized, as they disappear even for $p = 0.5$. They are also limited to a narrow voltage and magnetization angle regime; this is illustrated in Fig.(\ref{fig: 7}), which maps $PC_{\tau_{2},\tau_{1}}$ and $R$ as a function of both $V_{SD}$ and $\phi$.

At low voltages, in the Coulomb blockade regime, the second spin-channel is completely closed and the mechanism of another majority spin electron tunneling in is not allowed. At high voltages, however, the correlations also disappear because the dot can now easily be doubly occupied and the accumulated spin does not tend to negative values at long times. The narrow range of magnetization angles, on the other hand, is understood by considering what happens as the leads move away from parallel alignment. As $\phi \rightarrow \pi/2$, spin-precession allows the minority spin electron to couple to states in the drain and thus escape the dot at earlier times, and as $\phi \rightarrow \pi$, the current as a whole is diminishes, which directly suppresses the correlations as well.}


\begin{figure}
\centering
\includegraphics[width=0.7\columnwidth]{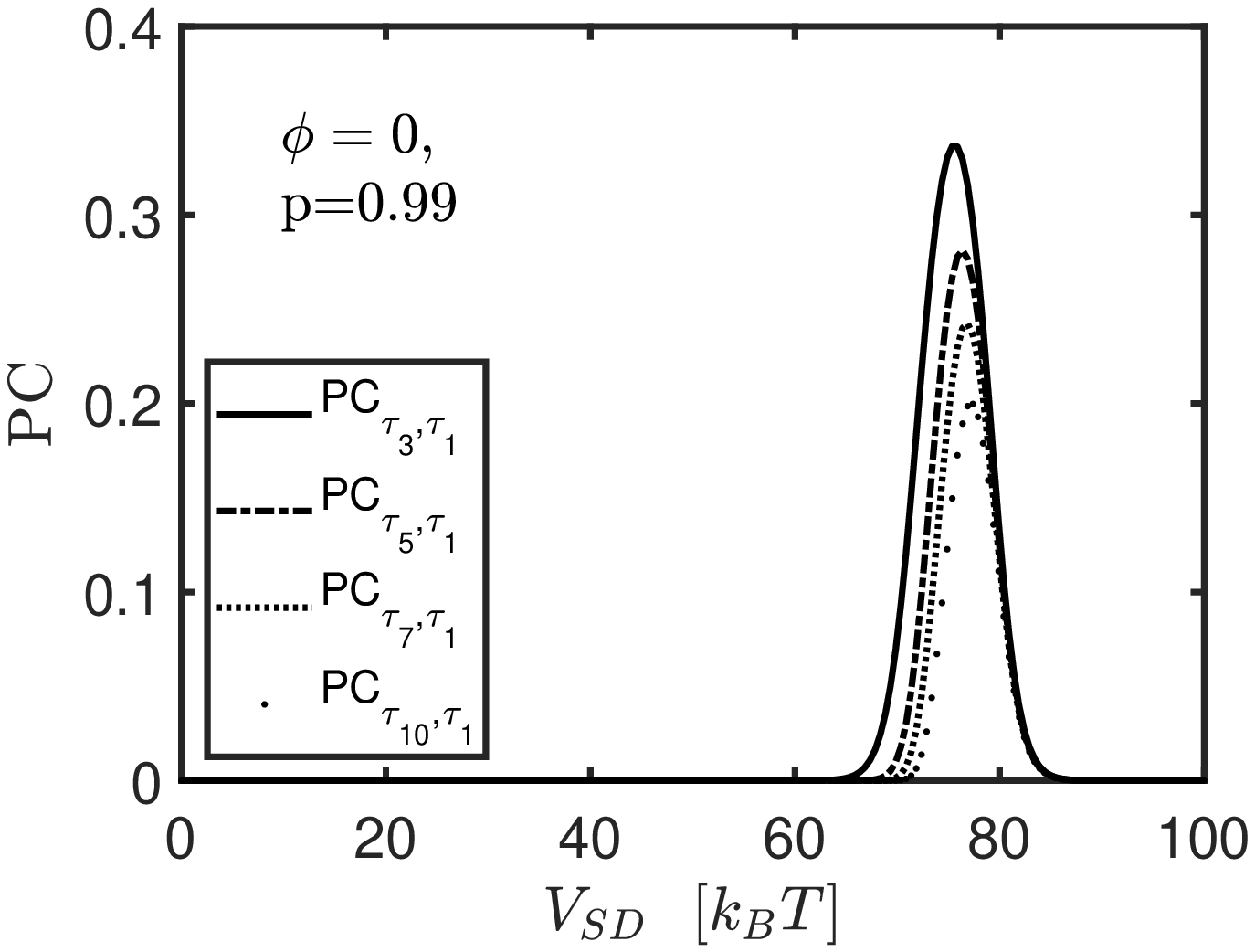}
\caption{Higher-order temporal correlations between $\tau_{1}$ and $\tau_{n}$, $PC_{\tau_{n},\tau_{1}}$, as a function of bias voltage $V_{SD}$. All parameters are the same as in  Figs.(\ref{PV_parallel}).}
\label{fig: higherorderPC}
\end{figure}

{   In an experimental setup, it is unlikely that one would have a completely efficient detector capable of measuring each electron tunneling event. A natural question, then, is whether these correlations survive into later tunnelings; that is, if the detector misses a series of waiting times after the first one, $\tau_{2},\dots,\tau_{n-1}$, but captures the $n^{\text{th}}$ waiting time, $\tau_{n}$, are $\tau_{1}$ and $\tau_{n}$ still correlated?

Fig.(\ref{fig: higherorderPC}) shows that, for this model, this is indeed the case. The initial waiting time is non-negligibly correlated with many successive waiting times, even up to the $w_{1}(\tau_{10})$, as shown in the plot. As one would expect, however, as $n$ increases, the Pearson coefficient between $\tau_{1}$ and $\tau_{n}$ decreases. The peak in each curve also occurs over a similar voltage range, indicating that the same processes discussed previously also produce the correlations in higher-order waiting times.}

\section{Conclusions}\label{sec: Conclusions}

Molecular spin-valves are experimentally realizable transport junctions that offer the possibility of harnessing the quantum spin, rather than just the electric charge, of tunneling electrons. In this paper, we explore the extent to which current fluctuations can be controlled in a spin-valve system, using the WTD and its associated statistics, alongside a Born-Markov master equation approach to describe the system dynamics.

First, we analyzed the randomness parameter, $R$, as a function of spin-polarization and relative magnetization angle, finding similar behavior to that of the Fano factor in the same regimes. For parallel magnetizations, and in the single occupancy voltage regime, $R \gg 1$ when $ p \rightarrow 1$; the small percentage of minority spin tunnelings cause severe electron bunching. This is also evident for perpendicular orientations, as the NDR in the current-voltage plot manifests as $R > 1$ too. Here, though, the level of electron bunching is much lower, as Larmor-like spin precession somewhat mitigates the effect. Sub-Poissonian transport is observed for the case of anti-parallel leads and low magnetization, in accordance with known behavior in an Anderson model, while high magnetization strongly suppresses the current and thus drives transport towards Poissonian behavior. Most importantly, it is observed that all three transport regimes are accessible simply by changing only the applied voltage and relative orientation of lead magnetization.

Temporal correlations are absent for any system arrangement allowing only a single transport channel, while voltages well above the double occupation threshold yield weak negative correlations. The most interesting region occurs within a small voltage band around the opening of the second transport channel and small angular deviation from the parallel orientation; strong temporal correlations emerge for highly magnetic leads within this region. From the scaled correlation map between all possible pairs of successive waiting times, one can see that, while two short waiting times are uncorrelated, two long waiting times are strongly positively correlated. For this to occur, minority spin accumulates in the molecule during the first long waiting time and is retained after the next tunneling event, making it highly likely that a second long waiting time follows the first. {   For this mechanism to function, the minority spin electron must become thoroughly trapped within the molecule; the leads must be highly spin-polarized to severely limit availability of minority electrons. This is assisted by a significant Coulomb repulsion, which allows trapped minority electrons to impede majority electrons. Increasing the voltage bias well into the double occupancy regime diminishes the effect by allowing electrons to easily overcome repulsion, while deviating from the parallel electrode orientation aids the mobility of minority spin electrons by means of greater Lamor-like precession. 

While the focus is largely on exploring correlations between successive waiting times, a surprising observation is made: significant correlations also extend to higher-order waiting time distributions. Driven by the same processes as those in successive waiting times, correlations between the $1^{\text{st}}$ and the $n^{th}$ waiting time decreases with $n$. These correlations are significant even up to tenth-order: the maximum value of $PC_{\tau_{10},\tau_{1}}=0.1998$.

Not only are the aforementioned temporal correlations relatively large, the various regimes are accessible within a single device by manipulation of both $V_{SD}$ and $\phi$, each over a short range. This demonstrates how the spin-valve device can controllably switch between regimes to provide versatile information transfer. It should be noted that these functions are highly sensitive to electron detection errors. Specifically, if the function of the device is to encode information in the duration of the waiting time, failing to detect a single electron will cause significant processing errors. However, the observation of notable correlations in higher-order waiting times does offer some robustness, as the third waiting time is still correlated with the first even if the second is undetected. 
}

\section{Acknowledgments}

The authors thank Julian Lawn for creating Figs.(\ref{fig:1})-(\ref{fig:2}) as well as his helpful discussions.

\appendix

\section{Spin-resolved self-energies}\label{app: Spin-resolved self-energies}

The self-energies in Eq.\eqref{eq: ME final} are Fourier transforms of bath-correlation functions, obtained using standard methods \cite{Breuer2002}, and can be written as
\begin{align}
\Sigma^{<}_{\alpha,\sigma\sigma'}(\omega) & = \Delta^{<}_{\alpha,\sigma\sigma'}(\omega) + \frac{i}{2}\gamma^{\alpha}_{\sigma\sigma'}f^{+}_{\alpha}(\omega) \\
\Sigma^{>}_{\alpha,\sigma\sigma'}(\omega) & = -\Delta^{>}_{\alpha,\sigma\sigma'}(\omega) - \frac{i}{2}\gamma^{\alpha}_{\sigma\sigma'}f^{-}_{\alpha}(\omega).
\end{align}
Here, $f^{\pm}_{\alpha}(\omega)$ are the relevant Fermi-Dirac functions for electrode $\alpha$,
\begin{align}
f^{\pm}_{\alpha}(\omega) & = \frac{1}{1 + e^{\pm\left(\omega - \mu_{\alpha}\right)/k_{B}T}},
\end{align}
and $\Delta^{<,>}_{\alpha,\sigma\sigma'}(\omega)$ and $\gamma^{\alpha}_{\sigma\sigma'}$ describe the renormalization and broadening of the system energy levels due to interaction with the electrodes, respectively. 

The energy-independent $\gamma^{\alpha}_{\sigma\sigma'}$ terms are
\begin{align}
\gamma^{S}_{\sigma\sigma} & = \Gamma_{S}\left(1 \pm p_{S}\right) \\
\gamma^{S}_{\sigma\bar{\sigma}} & = 0 \\
\gamma^{D}_{\sigma\sigma} & = \Gamma_{D}\left[1 \pm p_{D}\cos(\phi)\right] \\
\gamma^{D}_{\sigma\bar{\sigma}} & = \Gamma_{D}\sin(\phi)p_{D},
\end{align}
where $\uparrow$ and $\downarrow$ correspond to signs of $+$ and $-$, respectively. In the wideband limit, the renormalizations can be calculated analytically using the digamma function, $\psi(x)$ \cite{Rudge2020}:
\begin{align}
\Delta^{<}_{\alpha,\sigma\sigma'}(\omega) & = \frac{\gamma^{\alpha}_{\sigma\sigma'}}{2\pi} \Re \left\{\psi\left(\frac{1}{2} + \frac{i}{2\pi k_{B}T}\left(\omega - \mu_{\alpha}\right)\right)\right\},
\end{align}
with $\Delta^{>}_{\alpha,\sigma\sigma'}(\omega) = -\Delta^{<}_{\alpha,\sigma\sigma'}(\omega)$.

\section{Spin-valve BMME in Liouville space}\label{app: Spin-valve BMME in Liouville space}

To move from the GME in Hilbert space to a time-local differential equation in Liouville space, the system dynamics are collected into one superoperator, $\mathbf{L}$, which has the form
\begin{align}
    \mathbf{L}= \sum_\alpha ( \mathbf{A}_{\alpha} -i  \mathbf{D}_{\alpha}), \label{eq: Liouvillian defn}
\end{align}
where
\begin{widetext}
\begin{align}
\mathbf{A}_{\alpha} & = \left[\begin{array}{c c c c c c}
-2\Gamma_{\alpha}f_{\alpha}^{+} & \gamma^{\alpha}_{\uparrow \uparrow}f_{\alpha}^{-} & \gamma^{\alpha}_{\uparrow \downarrow}f_{\alpha}^{-} & \gamma^{\alpha}_{\downarrow \uparrow}f_{\alpha}^{-} & \gamma^{\alpha}_{\downarrow \downarrow}f_{\alpha}^{-} & 0 \\
\gamma^{\alpha}_{\uparrow \uparrow}f_{\alpha}^{+} & -\gamma^{\alpha}_{\downarrow \downarrow}f_{\alpha,U}^{+}-\gamma^{\alpha}_{\uparrow \uparrow}f_{\alpha}^{-} &\frac{1}{2}\gamma^{\alpha}_{\uparrow \downarrow}(f_{\alpha,U}^{+}-f_{\alpha}^{-}) & \frac{1}{2}\gamma^{\alpha}_{\uparrow \downarrow}(f_{\alpha,U}^{+}-f_{\alpha}^{-}) & 0 & \gamma^{\alpha}_{\downarrow \downarrow}f_{\alpha,U}^{-} \\
\gamma^{\alpha}_{\uparrow \downarrow}f_{\alpha}^{+} & \frac{1}{2}\gamma^{\alpha}_{\uparrow \downarrow}(f_{\alpha,U}^{+}-f_{\alpha}^{-}) & \Gamma_{\alpha}(-f_{\alpha,U}^{+}-f_{\alpha}^{-}) & 0 & \frac{1}{2}\gamma^{\alpha}_{\uparrow \downarrow}(f_{\alpha,U}^{+}-f_{\alpha}^{-}) & -\gamma^{\alpha}_{\uparrow \downarrow}f_{\alpha,U}^{-} \\
\gamma^{\alpha}_{\uparrow \downarrow}f_{\alpha}^{+} & \frac{1}{2}\gamma^{\alpha}_{\uparrow \downarrow}(f_{\alpha,U}^{+}-f_{\alpha}^{-}) & 0 &  \Gamma_{\alpha}(-f_{\alpha,U}^{+}-f_{\alpha}^{-})  & \frac{1}{2}\gamma^{\alpha}_{\uparrow \downarrow}(f_{\alpha,U}^{+}-f_{\alpha}^{-}) & -\gamma^{\alpha}_{\uparrow \downarrow}f_{\alpha,U}^{-} \\
\gamma^{\alpha}_{\downarrow \downarrow}f_{\alpha}^{+} & 0 & \frac{1}{2}\gamma^{\alpha}_{\uparrow \downarrow}(f_{\alpha,U}^{+}-f_{\alpha}^{-}) & \frac{1}{2}\gamma^{\alpha}_{\uparrow \downarrow}(f_{\alpha,U}^{+}-f_{\alpha}^{-}) & -\gamma^{\alpha}_{\uparrow \uparrow}f_{\alpha,U}^{+}-\gamma^{\alpha}_{\downarrow \downarrow}f_{\alpha}^{-} & \gamma^{\alpha}_{\uparrow \uparrow}f_{\alpha,U}^{-} \\
0 & \gamma^{\alpha}_{\downarrow \downarrow}f_{\alpha,U}^{+} & -\gamma^{\alpha}_{\uparrow \downarrow}f_{\alpha,U}^{+} & -\gamma^{\alpha}_{\uparrow \downarrow}f_{\alpha,U}^{+} &\gamma^{\alpha}_{\uparrow \uparrow}f_{\alpha,U}^{+} & -2\Gamma_{\alpha}f_{\alpha,U}^{-} \end{array}\right] \text{ and} \label{eq: matrix A}
\end{align}
\end{widetext}
\begin{align}
\mathbf{D}_{\alpha} &  = \left[\begin{array}{c c c c c c}
0 & 0 & 0 & 0 & 0 & 0\\
0 & 0 & -\gamma^{\alpha}_{\uparrow\downarrow}\delta_{\alpha} & -\gamma^{\alpha}_{\uparrow\downarrow}\delta_{\alpha} & 0 & 0  \\
0 &  \gamma^{\alpha}_{\uparrow\downarrow}\delta_{\alpha} & 0 & 0 & -\gamma^{\alpha}_{\uparrow\downarrow}\delta_{\alpha} & 0 \\
0 & -\gamma^{\alpha}_{\uparrow\downarrow}\delta_{\alpha} &  0 & 0 & \gamma^{\alpha}_{\uparrow\downarrow}\delta_{\alpha} & 0 \\
0 & 0 & -\gamma^{\alpha}_{\uparrow\downarrow}\delta_{\alpha} & -\gamma^{\alpha}_{\uparrow\downarrow}\delta_{\alpha} & 0 & 0  \\
0 & 0 & 0 & 0 & 0 & 0 
\end{array}\right]. \label{eq: matrix D}
\end{align}

\newpage

In defining the matrices in Eq.\eqref{eq: matrix A} and Eq.\eqref{eq: matrix D}, we have introduced the new notation
\begin{align}
\delta_{\alpha} & = \frac{1}{2\pi}\Re\left\{\psi\left(\frac{1}{2}+\frac{i}{2\pi k_{B}T}(\varepsilon+U-\mu_{\alpha})\right) - \right.\nonumber \\
& \hspace{1.5cm} \left.\psi\left(\frac{1}{2}+\frac{i}{2\pi k_{B}T}(\varepsilon-\mu_{\alpha})\right)\right\}, \\
f_{\alpha}^\pm & = f_{\alpha}^\pm(\varepsilon), \hspace{0.5cm} f_{\alpha U}^\pm = f_{\alpha}^\pm(\varepsilon+U).
\end{align}

Following the procedure outlined at the start of Section \ref{subsec: WTD}, the jump superoperator is defined as
\begin{align}
    \mathbf{J} & = \left[\begin{array}{c c c c c c}
0 & \gamma^{D}_{\uparrow \uparrow}f_{D}^{-} & \gamma^{D}_{\uparrow \downarrow}f_{D}^{-} & \gamma^{D}_{\uparrow \downarrow}f_{D}^{-} & \gamma^{D}_{\downarrow \downarrow}f_{D}^{-} & 0 \\
0 & 0 & 0& 0 & 0 &  \gamma^{D}_{\downarrow \downarrow}f_{D,U}^{-} \\
0 & 0 & 0& 0  & 0 & -\gamma^{D}_{\uparrow \downarrow}f_{D,U}^{-} \\
0 & 0 & 0 &0 & 0 & -\gamma^{D}_{\uparrow \downarrow}f_{D,U}^{-}\\
0 & 0 & 0 & 0 & 0 & \gamma^{D}_{\uparrow \uparrow}f_{D,U}^{-}\\
0 & 0 & 0 & 0 & 0 & 0
\end{array}\right].
\end{align}

\bibliography{Main_text_incl._fig}

\end{document}